\documentclass{ws-procs9x6}
\usepackage{astron}

\def\apj{ApJ}                 % Astrophysical Journal
\def\apjl{ApJ}                % Astrophysical Journal, Letters
\def\aap{A\&A}                % Astronomy and Astrophysics
\def\mnras{MNRAS}             % Monthly Notices of the RAS
\begin{document}

\title{LOFAR: A new radio telescope for low frequency radio
observations: \\ {\it Science and project status}}

\author{H. R\"ottgering}

\address{Sterrewacht Leiden,
P.O. Box 9513, 
2300 RA Leiden, 
The Netherlands \\
E-mail: rottgeri@strw.leidenuniv.nl}

\author{A. G. de Bruyn, R. P. Fender, J. Kuijpers} 

\address{ASTRON/Groningen, Amsterdam, Nijmegen, The Netherlands}

\author{M. P. van Haarlem, M. Johnston-Hollitt, G. K. Miley}

\address{ASTRON, Leiden, Leiden, The Netherlands}


\maketitle

\abstracts{LOFAR, the Low Frequency Array, is a large radio telescope
consisting about 100 soccer field sized antenna stations spread over a
region of 400 km in diameter. It will operate in the frequency range
from $\sim 10$ to 240 MHz, with a resolution at 240 MHz of better than
an arcsecond. Its superb sensitivity will allow for a broad range of
astrophysical studies. In this contribution we first discuss four
major areas of astrophysical research in which LOFAR will undoubtedly
make important contributions: reionisation, distant galaxies and AGNs,
transient radio sources and cosmic rays. Subsequently, we will discuss
the technical concept of the instrument and the status of the LOFAR
project.}

\section{Introduction}

LOFAR, the Low Frequency  Array, is a large radio telescope that
will open up the virgin territory of observations at low radio
frequencies for a broad range of astrophysical studies.  It will
observe the Universe at frequencies from $\sim$ 10 to 240 MHz
(corresponding to wavelengths of 30 to 1.5 m) with a resolution at
240 MHz of better than one arcsecond.  Its superb sensitivity and
high resolution will be a dramatic improvement over previous
facilities at these wavelengths. It is only with recent
developments in computer hardware, software and
broad-band internet connectivity, that the construction of this
telescope, the calibration of the associated data 
and generation of high fidelity, high resolution wide-field images has
become possible.

\section{Science}

The science case for LOFAR is broad and has several interesting
applications outside the field of extra-solar astrophysics, such as
studies of the Earth's ionosphere and the physical properties of
the solar wind. For astrophysical research a number key scientific
areas have been identified, which include:

\begin{itemize}
\item the sources and epoch of reionisation;
\item the formation and evolution of galaxies and AGN;
\item the nature of transient sources and high energy objects, and
\item the origin of high energy cosmic rays.
\end{itemize}

We will discuss these in turn.

\subsection{Epoch of reionisation}

One of the most exciting goals of LOFAR will be to chart the end
of the ``Dark Ages'' when the first stars and AGNs started to ionise
the neutral baryonic gas pervading the Universe. In the joint
discussion at this Texas symposium Madau has extensively discussed
the theoretical models and observational status of this ``epoch of
reionisation''.

LOFAR's ability to search for and study the redshifted 21cm
emission line at the redshift range of $z \sim 5-15$ will open up
a window onto (literally) one of the most exciting periods in
cosmic history. Furthermore, LOFAR will be able to carry out these studies with
an angular resolution an order of magnitude better than WMAP.
Towards the end of this decade the JWST and ALMA will begin
directed studies of individual objects beyond $z=7$. However, the
field of view will be very small, and the number of observable
objects in or before the Epoch of Reionisation (EoR) will be few.
This will limit the usefulness of these instruments for the study of the
large-scale distribution of HI and HII, `the raw stuff from which
stars and galaxies are made'. Topics that LOFAR will address
include:

\begin{itemize}
\item The history of reionisation, i.e. the redshift range in which
the bulk of the HI became ionized. Identification of possible
different stages of the process.

\item The spatial distribution of ionised and neutral IGM, and its evolution
during the epoch of reionisation.

\item The objects responsible for reionising the Universe.
Proto-galaxies and their massive stellar populations are the most
likely sources, but the role of the first generation of quasars
remains unclear.
\end{itemize}

\subsection{Galaxy formation and evolution}

One of the most intriguing problems in modern astrophysics
concerns the formation of massive black holes, galaxies and
clusters of galaxies. There are three main classes of objects in
the early Universe that will be observed by LOFAR with the goal of
investigating questions related to the formation of these objects.
These key types of objects are (i) distant radio sources, produced
by black holes in the nuclei of massive galaxies, (ii)
``starburst'' galaxies, i.e. infant galaxies observed to be
undergoing a vigorous episode of star formation and (iii) diffuse
radio emission as probes of gas in clusters of galaxies.

The most efficient method for finding {\it distant radio galaxies} uses an
empirical correlation between radio spectral steepness and distance
(e.g. de Breuck et al. 2000). \nocite{bre00a} 
Using this relation, 
LOFAR will efficiently pick
out radio galaxies at larger distances than currently possible.
Study of these distant radio galaxies at other
wavelengths will provide information about the formation of massive
galaxies and AGN. Furthermore, since distant radio galaxies pinpoint
proto-clusters, studying the environment of these distant galaxies
will constrain the formation of clusters at the earliest epochs
(e.g. Venemans et al. 2002). \nocite{ven02} It is possible that some of these radio
galaxies are located at an epoch before reionisation has completely
occurred. This would open up the possibility of studying the epoch of
reionisation through observations of the absorbing neutral gas against
these very distant radio galaxies (Carilli et al. 2002)
\nocite{car02b}

With its unprecedented sensitivity to non-thermal radio emission from
star formation, LOFAR will detect large numbers of {\it star-forming
galaxies} at an epoch at which the bulk of galaxy formation is believed
to occur.  Since the ratio of radio flux to sub-mm flux is a sensitive
redshift indicator (Carilli and Yun 1999), \nocite{car99} LOFAR
surveys, in combination with data from new far-IR and millimeter
facilities such as SIRTF, ALMA, and JWST, will provide
distances and thus allow for a complete census of the cosmic
star-formation history, unhindered by the effects of dust obscuration.

{\it Clusters of galaxies} often contain diffuse radio sources that
are shaped by the dynamics of the gas in which they are embedded,
LOFAR will be able to detect and study these radio
sources in the many tens of thousands of clusters up to redshifts of
two that will be detected using the XMM X-ray telescope, the Planck
satellite, and the Sloan Digital Sky Survey (e.g. En{\ss}lin,
T.~A. and R{\" o}ttgering, 2002). \nocite{ens02b} Such studies will
be very relevant for (i) understanding the dynamics of the cluster gas
(ii) determining the origin of their magnetic field content, and (iii)
constraining physical models for the, as yet unknown,  origin of these sources.

\subsection{The bursting and transient universe}

LOFAR's large instantaneous beam, will make it  uniquely suited to
efficiently monitor a large fraction of the sky, allowing for the
first time a sensitive unbiased survey for radio transients on a
variety of time scales, ranging from a few tenths of seconds to many
days. Rapid follow up with LOFAR at high resolution will provide accurate
positions required for optical and X-ray identifications. Table
1 gives  an overview of the classes of object known or
expected to exhibit variable radio emission. Also indicated are
the variability time-scales, the number of objects/events expected
to be observed per year and an estimate of the distances to which
these objects can be seen.

\begin{table}[h]
\tbl{Overview of transients expected to be detected and monitored
with LOFAR. Also indicated are the variability time-scales, the
number of objects/events expected to be observed per year and an
estimate of the distances to which these objects can be seen.}
 {
\small
 \begin{tabular}{|l|c|c|c|}
        \hline \hline
  Object & Variability& No. of& Maximum\\
  & Timescale&Events&Distance\\
          \hline \hline
  Radio Supernovae& days--months& $\sim 3$/yr & 2--3 $\times$ Virgo Cluster\\
  \hline
  GRB Afterglows& days--months& $\sim 100$/yr & Observable Universe \\
  \hline
  Galactic Black Holes& days--months& 10--100/yr & Local Group\\
  and Neutron Stars&&&\\
  \hline
  Pulsars& millisec--sec& few 1000& Whole Galaxy, M31\\
  \hline
  Intermediate mass& days$?$& 1--5/yr& Virgo Cluster\\
  Black Holes&&&\\
  \hline
  Exoplanets&minutes--hours&10--100& 30 pc\\
  \hline
  Flare Stars& millisec--hours& 100--1000& $<$ 1 kpc\\
  \hline
  `LIGO Events'& $\leq$millisec & few$?$& Observable Universe\\
          \hline \hline
 \end{tabular}
}
\end{table}

As can be seen from Table 1, for high energy astrophysics there are a
number of particular interesting applications. From the empirical
relation between radio and X-ray emission for Gamma-ray bursters and
Galactic black-hole/neutron-star it is clear that the all-sky
monitoring with LOFAR will be a factor of 5--10 more effective in
discovering such events than previous all-sky-monitors.  It is
therefore anticipated that LOFAR will be the primary source of
triggers for the high-energy community utilising target-of-opportunity
programs on e.g. HST / VLT / Chandra / XMM etc. Furthermore, several
models for strong `LIGO events', for example the coalescence of two
neutron stars, are predicted to have an associated strong burst of
radio emission (e.g. Hansen and Lyutikov 2000). \nocite{han01}
Similarly, prompt radio emission is predicted by some GRB models.

\subsection{LOFAR as a cosmic ray detector}

The existence of high-energy cosmic rays (HECRs) at energies
between 10$^{15}$ -- 10$^{20.5}$ eV is an outstanding challenge
for particle astrophysics. Both the sites and processes for
accelerating particles are unknown. Possible candidate sources of
these HECRs are shocks in radio lobes of powerful radio galaxies,
intergalactic shocks created during the epoch of galaxy formation,
so-called Hyper-novae, Gamma-ray bursts and magnetars.
Alternatively, HECRs are perhaps decay products of super-massive
particles from topological defects, left over from phase
transitions in the early universe.

A primary CR induces a particle cascade in the atmosphere which is
aligned along the direction of motion of the primary particle. A
substantial part of such a ``CR shower'' is leptonic and produces
coherent radio emission (e.g. Cherenkov emission, transition
radiation, and synchrotron emission) in the terrestrial
magnetosphere.  At the high Lorentz factors considered here, the
cascade is confined to a slab a few meters wide 
perpendicular to the travel direction, which emits coherent
`geo-synchrotron' radiation below 200 MHz (e.g. Falcke and 
Gorham 2002). \nocite{fal02} From the arrival times
and intensities of the radio pulse at various antennas of LOFAR,
the direction of the primary particle can then be determined to an
accuracy of 1 degree.

LOFAR has a unique potential for studies of HECRs, including:
\begin{itemize}
\item The study of HECRs in the wide energy range from $10^{15}$ eV
to $10^{20.5}$ eV with one and the same instrument; The  standard
optical methods (Cherenkov or fluorescence) which  have only a
10\% duty cycle, observe only a relative narrow range of particle
energies;
\item Investigating the poorly understood development of the electromagnetic
part of the cascade by {\it in situ} radio observations;
measurement of the height of the shower maximum, the forward
cross-sections and inelasticity parameter in high energy particle
collisions which cannot be determined in a particle collider.
\item The discovery of point-sources in high-energy ($>10^{18}$ eV)
neutrons which can cross the Galaxy before they decay, and whose
observation may thus reveal their origin.  Discrimination between
anisotropies caused by charged nuclei whose paths are affected by
the Galactic magnetic field and neutrons is, in principle,
possible by studying the absence or presence of such anisotropies
at lower energies;
\item Measurement of the composition of HECRs from the study of
simultaneous pairs of showers at a distance up to several 100 km.
Such `multiplet' events are expected from photodisintegration of
CRs in the solar radiation field (Gerasimova-Zatsepin effect);
\item Detection of coherent radio emission from neutrinos at energies
$10^{15}$ -- $10^{18}$ eV in horizontal showers, and of tau
neutrino events (`double-bang': two showers at 50 km distance).
\end{itemize}

\begin{figure}[t]
\centerline{\psfig{file=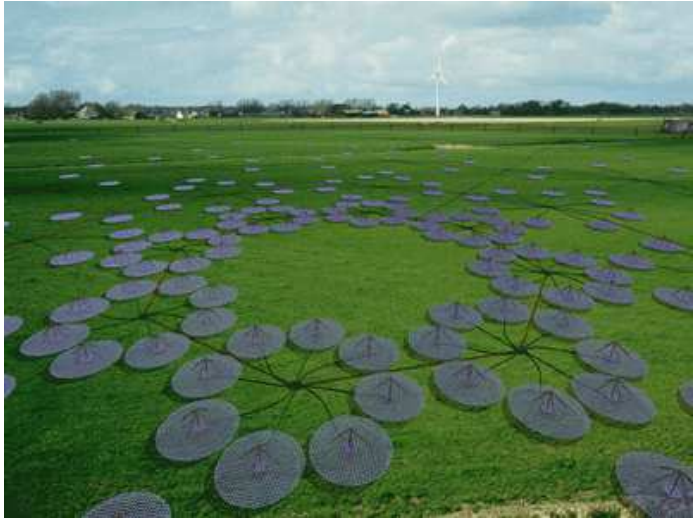,width=0.8\textwidth}}
\caption{\label{ant} An artist impression of the candidate
layout of (a Netherlands version of) an individual LOFAR station.}
\end{figure}

\section{The telescope}

LOFAR will have 2 antenna systems, one for the 10-90 MHz range,
and one for the 110-240 MHz range. The antennas will be placed in
soccer-field sized stations yielding, for each station, effective
apertures that will range from 50\, m to 150\, m, depending on
frequency. For an artist impression of (the Netherlands version
of) such a station see Fig. \ref{ant}.

The signals from each antenna are digitised and  fed into the
station beamformer. The beamformer can produce up to eight
coherent ``station-beams'' within the primary power pattern of the
antenna element in use.  The output data stream from each station
beam is sent to an optical link for transmission to the central
processing facility.

\begin{figure}[t]
\centerline{\psfig{figure=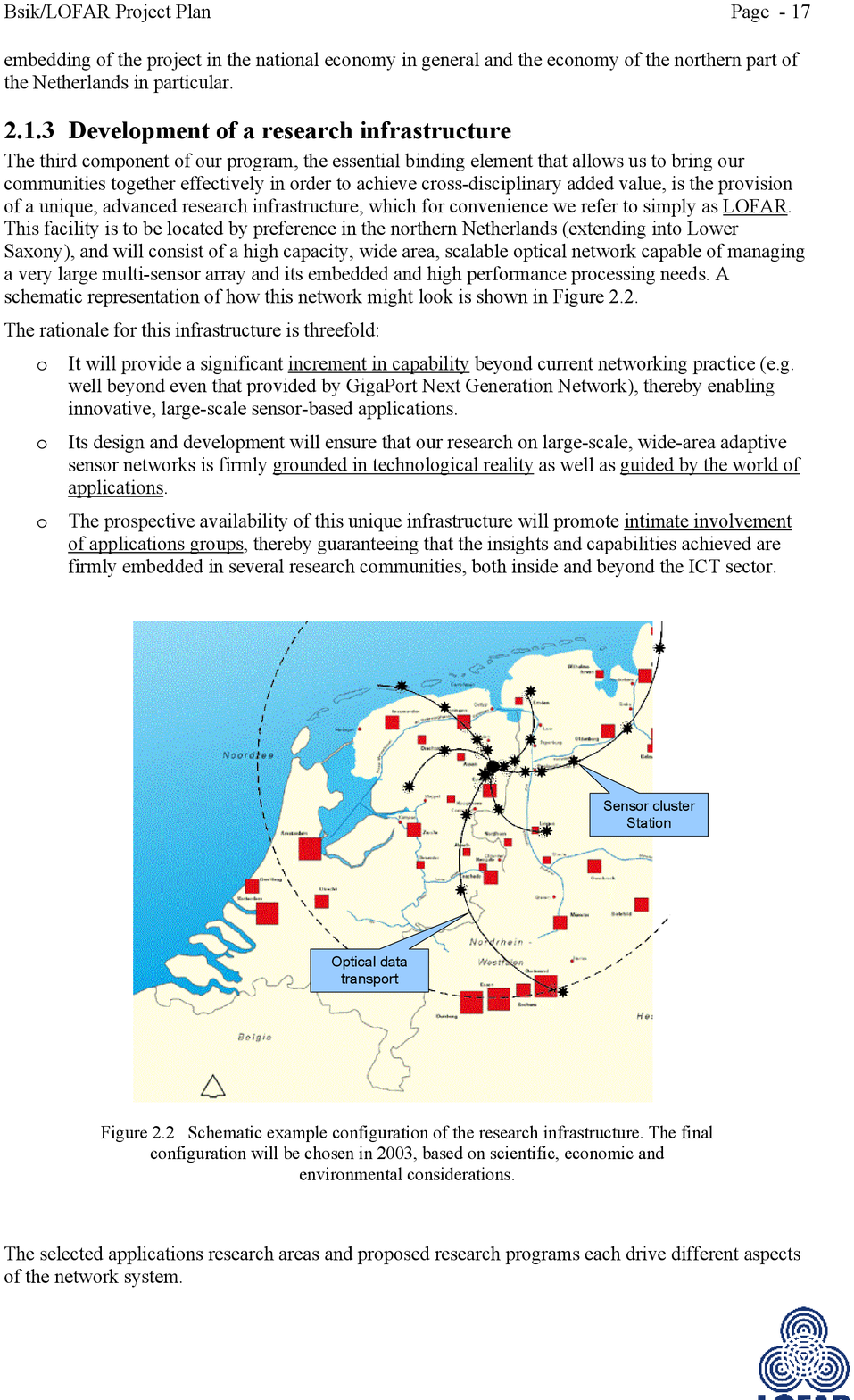,width=0.8\textwidth,clip=}}
\caption{\label{layout} The approximate layout of the LOFAR
stations in a 6-armed log-spiral pattern for the proposed Dutch site. 
(two other sites are also under consideration, for further info see text). 
}
\end{figure}

In total, of the order of 100 stations will make up the array.
These stations will be distributed over an area with a diameter of
about 400\, km (see Fig. \ref{layout}). This yields a maximum
angular resolution of about 0.6 arcseconds at the highest LOFAR
frequencies.

An iterative scheme based on existing successful self-calibration
techniques for radio astronomy will be used to calibrate LOFAR data.
This scheme, which is currently under development, will solve for the
ionospherically induced phase fluctuations, the characteristics of the
station beams and maps of the radio sky.  Furthermore, the removal of
radio frequency interference is an important issue.  This is
facilitated by the high spectral resolution of LOFAR for which only a
few percent of the spectrum is affected by RFI.  Finally, the foreseen
enormous data rates of maximally 25 Tb/s are a major
challenge. However, this makes LOFAR an interesting testbed for ICT
research and development, which in turn is helping to fund the
instrument.  \nopagebreak
\section{The Project}

ASTRON (Dwingeloo, the Netherlands), M.I.T. (Cambridge, USA) and the  Naval
Research Lab (Washington, USA) are responsible for the design,
construction, operation and software of the LOFAR telescope. The
schedule as agreed by the international LOFAR partners for the design
and construction of LOFAR is given in Table 2.

\begin{table}[h!]
\tbl{Schedule for the entire LOFAR project} {\small
\begin{tabular}{|l|l|}
        \hline \hline
Year & Milestone  \\
        \hline \hline
2003 & Integrated test station with 100 antenna \\
2005   & LOFAR core operational  \\
2006& Initial operations of central core plus first outer stations \\
2008 & Full operation \\
\hline \hline
\end{tabular}
}
\end{table}
A site characterisation committee is presently obtaining data
needed to assess the suitability of LOFAR siting in the
Netherlands, south-west Australia and the southern USA (Texas and New
Mexico).  A decision on the location for LOFAR is foreseen to be taken
in 2003.

The international project is supervised by an International
Steering Committee (ISC), consisting of directors of the
participating institutes. An Engineering Consortium (EC)
is responsible for the design and implementation of the instrument. 
The Science Consortium Board (SCB) is responsible for developing
the science case for LOFAR in close collaboration with the
community and gives scientific input to the EC.

The Science Consortium Board very much welcomes suggestions for
improving or optimising the design of the instrument for general or
very specific applications.  For further information, visit: www.LOFAR.org

\end{document}